\documentclass[aps,prb,reprint,superscriptaddress]{revtex4-2}

\usepackage{graphicx}
\usepackage{amsmath}
\usepackage{amssymb}
\usepackage{float}
\usepackage{bigints}
\usepackage[normalem]{ulem}
\usepackage[colorlinks,urlcolor=blue]{hyperref}
\usepackage[capitalise]{cleveref}
\usepackage{xcolor}

\begin{document}
    \title{Shot noise as a diagnostic in the $\nu=2/3$ fractional quantum Hall edge zoo}

	\author{Sourav Manna}
 \email{sourav.manna@weizmann.ac.il}
	\affiliation{Department of Condensed Matter Physics, Weizmann Institute of Science, Rehovot 7610001, Israel}
	\affiliation{Raymond and Beverly Sackler School of Physics and Astronomy, Tel-Aviv University, Tel Aviv, 6997801, Israel}
	
	\author{Ankur Das}
         \email{ankur@labs.iisertirupati.ac.in}
	\affiliation{Department of Condensed Matter Physics, Weizmann Institute of Science, Rehovot 7610001, Israel}
        \affiliation{Department of Physics, Indian Institute of Science Education and Research (IISER) Tirupati, Tirupati 517619, India}

 	\author{Yuval Gefen}
	\affiliation{Department of Condensed Matter Physics, Weizmann Institute of Science, Rehovot 7610001, Israel}

	\author{Moshe Goldstein}
	\affiliation{Raymond and Beverly Sackler School of Physics and Astronomy, Tel-Aviv University, Tel Aviv, 6997801, Israel}
 
\begin{abstract}
The $\nu = 2/3$ filling is the simplest paradigmatic example of a fractional quantum Hall state, which contains counter-propagating edge modes. These modes can be either in the unequilibrated regime or equilibrated to different extents,
on top of a possible edge reconstruction.
In the unequilibrated regime,
two distinct renormalization group fixed points have been previously proposed, namely Kane-Fischer-Polchinski and Wang-Meir-Gefen.
In the equilibration regime, different degree of thermal equilibration
may occur, while charge is fully equilibrated.
Here, we show that this rich variety of models can give rise to 
three possible conductance plateaus at 
$e^2/2h$ (recently observed in experiments), $5e^2/9h$ (predicted here), and $e^2/3h$ (observed earlier in experiments) in a quantum point contact geometry.
We identify different mechanisms for \emph{electrical shot noise} 
generation at these plateaus, which provides an experimentally accessible venue 
for distinguishing among the distinct models.
\end{abstract}	
\maketitle

\textit{The tragic and barbaric killing of Prof.\ Sergei Gredeskul and his wife, Viktoria, as they spent a quiet weekend at home on the accursed October 7th, 2023, was one of many tragedies that happened on that day and the following
months. A student of Ilya Lifshitz, and a scientific grandson of Lev Landau, Sergei left a great career at Kharkiv University, Ukraine, and moved in 1991 to Israel,
where he became a member of the Physics Department at Ben Gurion University. Sergei made numerous contributions to condensed matter physics; a common theme that can be found 
in many of his works is disorder and inhomogeneity. Even those who did not know him closely felt that Sergei 
deserved a scientific commemoration, and we are proud to be part of this endeavor.}

\section{Introduction}

The fractional quantum Hall (FQH) effect \cite{PhysRevLett.48.1559,PhysRevLett.50.1395} serves as an important 
platform for studying topologically ordered phases
of matter. 
This remarkable phenomenon manifests itself in a gapped bulk 
with gapless chiral edges, realizing the bulk-boundary correspondence.
The edge modes 
can carry both charge and energy. They may either be co-propagating
or counter-propagating. Interaction and disorder may give rise to 
emergent edge renormalization (for the case of counter-propagating modes), with the $\nu=2/3$
state serving as the simplest example \cite{Macdonald1990,Kane1994}.
The early proposal by MacDonald postulated
the edge modes to be $e$ and $e/3$ charge modes \cite{Macdonald1990,Wen1990,McD1991}. However, this proposed edge structure failed to provide a picture 
consistent with the experimental observations, namely 
two-terminal electrical conductance of $2e^2/3h$, and 
the failure to detect a
counter-propagating $e/3$ charge
mode \cite{West1992}.

Eventually, it was proposed that random disorder induced charge tunneling between the counter-propagating modes may play a crucial role \cite{Kane1994}. It was shown that,
at zero temperature, the system approaches
a disorder dominated 
coherent renormalization group (RG)
fixed point, known as the Kane-Fischer-Polchinski (KFP) RG fixed point \cite{Kane1994}.
At this
fixed point the edge consists of 
a $2e/3$ charge mode counter-propagating to a neutral mode. This edge structure is 
consistent with the experimentally
observed two-terminal electrical conductance of $2e^2/3h$.

\begin{figure}[!t]
	\includegraphics[width=0.99\columnwidth]{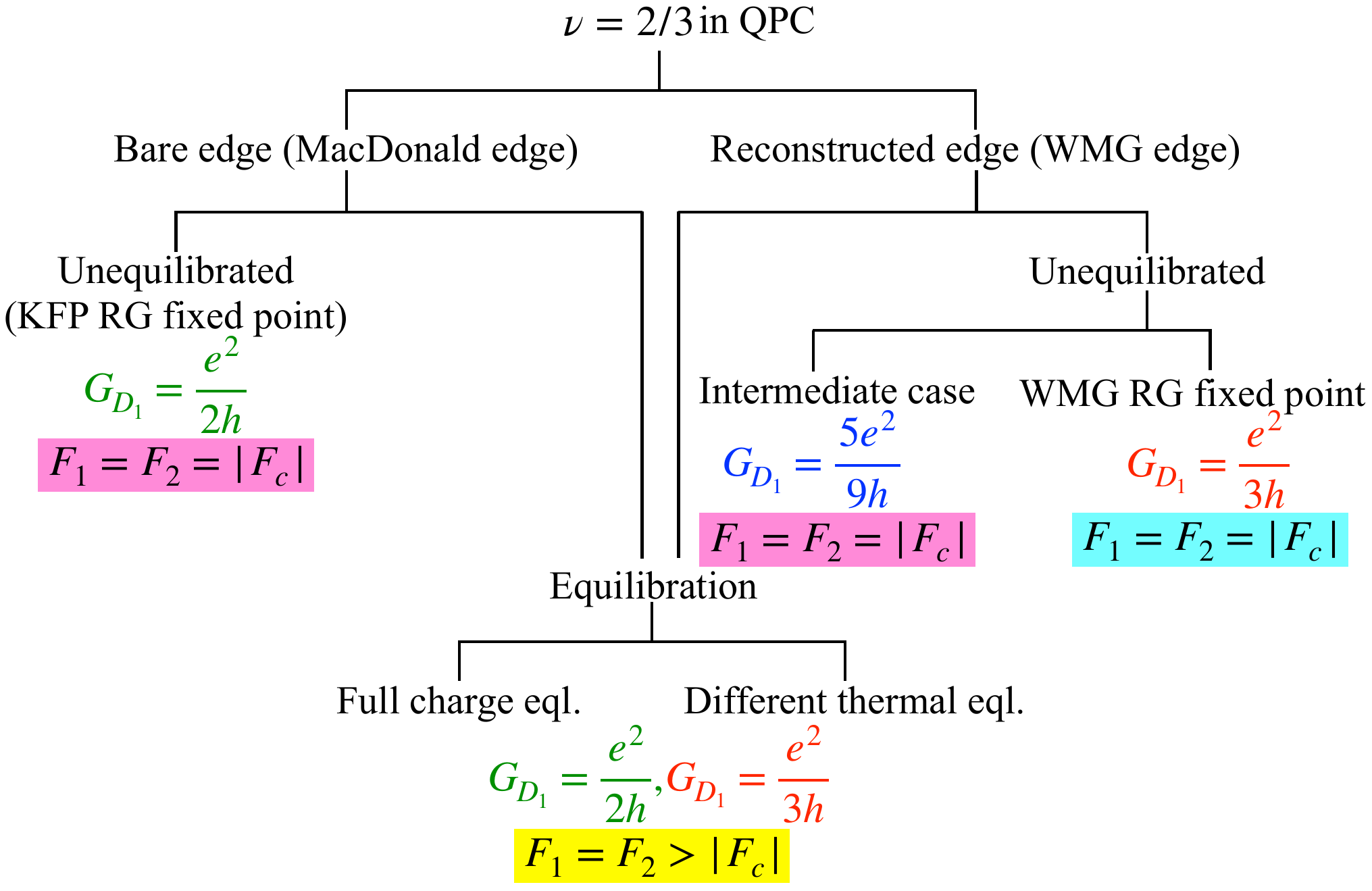}
	\caption{Color-coded conductance
 plateaus $(G_{D_1})$ and shot noise Fano factor $(F_1, F_2, F_c)$ inequalities: $e^2/2h$ QPC
 plateau (green) arises from either unequilibrated or equilibrated regimes, which are distinguishable by distinct Fano factor inequalities (pink and yellow), and similarly for the $e^2/3h$ QPC
 plateau. The $5e^2/9h$ QPC plateau originates only in the unequilibrated regime.}\label{FlowChart}
\end{figure}

A somewhat more complicated experimental device which can provide more information regarding the edge structure is a quantum point contact (QPC), a constriction in the two-dimensional electron gas \cite{PhysRevLett.75.3340}. Across the QPC, shot noise can be measured to 
determine the charge carried by an edge mode \cite{PhysRevB.45.1742,MartinNoiseReview,Mahalu1997,PhysRevLett.79.2526,Mahalu1998,Saminadayar1998,Dima2023}. 
For $\nu=2/3$, the observation of a 
$e^2/3h$ QPC conductance plateau
\cite{Cunningham1992,Mahalu2009}, measurements consistent with the existence of neutral modes \cite{Bid2010},
and a crossover of the effective charge while changing the temperature \cite{Mahalu2009}
were found.
These are incompatible with the KFP RG fixed point. To accommodate these experimental 
findings, 
reconstruction of the MacDonald edge was proposed leading to a
new coherent intermediate RG fixed point, proposed by Wang-Meir-Gefen (WMG) \cite{Wang2013}. At this
fixed point the edge consists of 
two $e/3$ charge modes counter-propagating to two neutral modes.

The existence of a plateau at a fractional QPC transmission (corresponding to a QPC filling) and shot noise therein can also be a consequence of
equilibration among the chiral edge modes \cite{PhysRevB.101.075308}
at finite temperatures.
Charge and heat equilibrations can occur independently.
Recently, different experiments 
\cite{PhysRevLett.126.216803, Melcer2022, Kumar2022, Srivastav2022} have confirmed that the thermal equilibration length
is order of magnitude larger than the charge equilibration length.
These findings have complicated the study of the steady state
of edges since different regimes of thermal equilibration
can occur while charge is fully equilibrated.

\begin{figure}[!t]
	\includegraphics[width=0.95\columnwidth]{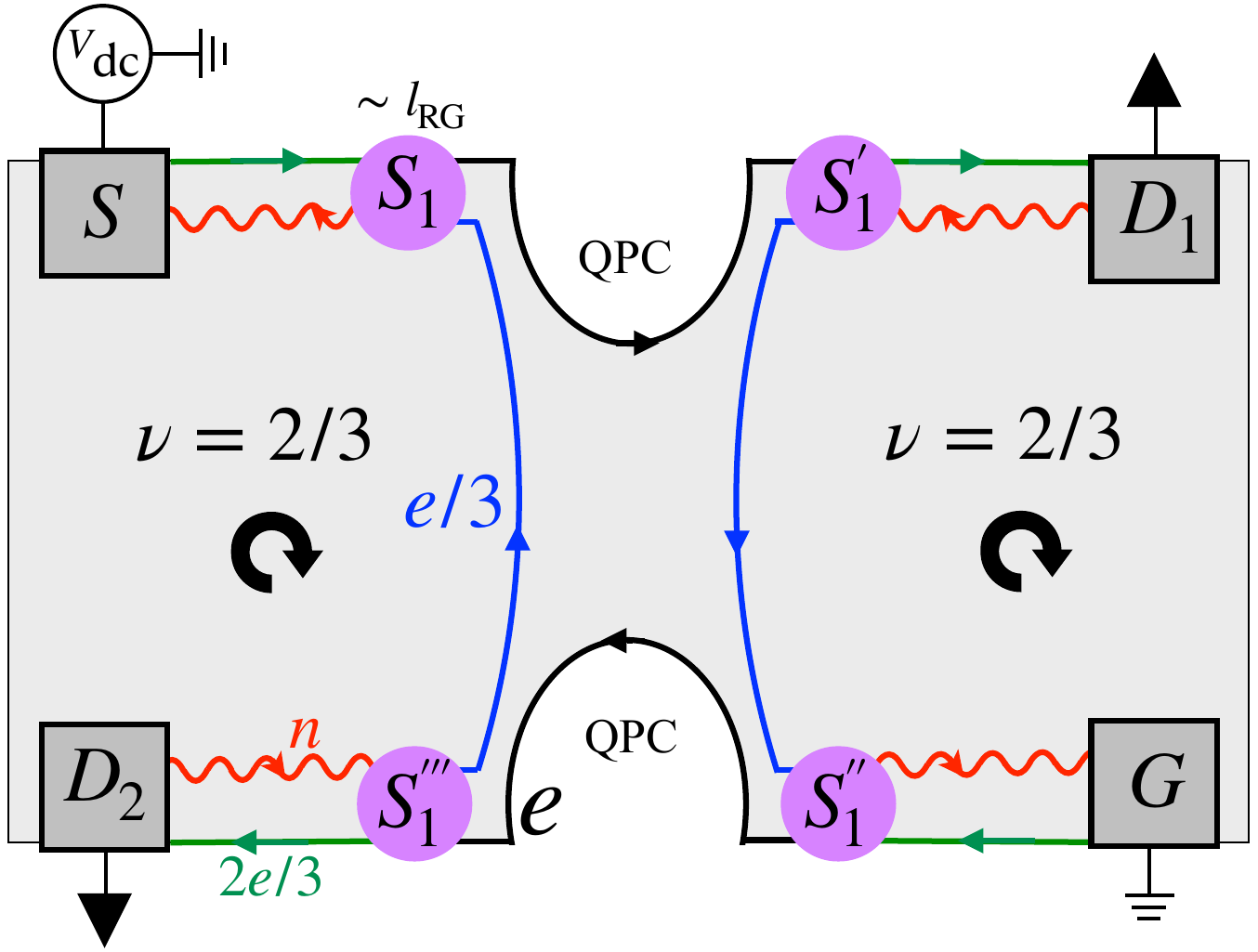}
	\caption{$1/2$ QPC conductance plateau 
 in the stochastic case: A Hall bar with bulk
 filling $\nu=2/3$ in a QPC. The device contains 
 a source $S$ (with dc voltage $V_{\text{dc}}$), a ground contact $G$, and two drains, $D_1$ and $D_2$. From the contacts
 emanate renormalized modes. The charge propagation
 chirality is depicted by a circular arrow.
 The scattering regions (purple circles) with size $l_{\text{RG}}$ are defined by the density kernel matrices $S_1,S_1',S_1'', S_1'''$ \cite{PhysRevB.52.R17040}. Here
$S_1'$ and $S_1'''$ account for the stochastic processes
and $S_1$ and $S_1''$ give rise to deterministic charge propagation.}\label{KFP_Fig}
\end{figure}

In this paper we show in details
(\cite{SMshort} contains the non-technical insight)
that these zoo of models 
can give rise to three distinct QPC conductance plateaus
at $e^2/2h, 5e^2/9h,$ and $e^2/3h$. We note that the $e^2/2h$ plateau has been reported 
recently \cite{Manfra2022,Hirayama2023} and the 
$e^2/3h$ plateau was discovered earlier \cite{Mahalu2009} 
in experiments. We predict the possible appearance of 
$5e^2/9h$ plateau. We identify different scenarios for each plateau, as well as different mechanisms which may give rise to \emph{electrical shot noise} at these plateaus. We show how shot noise can be used to experimentally discern between the different
models (we refer to \cref{FlowChart} 
for a flowchart, summarizing our findings). The rest of this paper is structured as follows:
We explain the emergence of $e^2/2h, 5e^2/9h,$ and $e^2/3h$
QPC conductance plateaus and calculate the shot noise for each
in \cref{1by2QPC_PRB}, \cref{5by9QPC_PRB}, and \cref{1by3QPC_PRB},
respectively. We provide a summary and an outlook in \cref{Summ}.

\section{The noisy $e^2/2h$ QPC plateau}\label{1by2QPC_PRB}

The recently observed $e^2/2h$ QPC conductance plateau \cite{Manfra2022,Hirayama2023} is analyzed in this
section (see also Refs.\ \onlinecite{PhysRevB.87.125130, Wang2021, Yuli2023}). We derive different inequalities
among autocorrelations and crosscorrelation 
in the unequilibrated and equilibrated regimes.

\subsection{Unequilibrated scenario}\label{KFP_stoch}

We briefly mention that in the fully coherent 
case, two distinct scenarios may occur:
either no QPC conductance plateau, or a noiseless
QPC plateau at the same conductance as we find below \cite{SMshort}.
We therefore assume that the wavepackets undergo incoherent scattering in
each region of size $l_{\text{RG}}$ (\cref{KFP_Fig}).
Density kernel matrices $S_1,S'_1,S''_1,S'''_1$ \cite{PhysRevB.52.R17040} quantify such processes. 
We note that $S_1'$ and $S_1'''$ give rise to the stochastic processes,
while $S_1$ and $S_1''$ give rise to deterministic processes.
Hence, we write
\begin{equation}
\begin{split}
&S_1=
\begin{pmatrix}
T_{11}  & R_{21} \\
R_{12} &  T_{22} 
\end{pmatrix}
=S_1'',\\&
S_1'=
\begin{pmatrix}
\langle T'_{11} \rangle & \langle R'_{21} \rangle\\
\langle R'_{12} \rangle & \langle T'_{22} \rangle 
\end{pmatrix}
=S_1''',
\end{split}
\end{equation}
where $T',T''',R',R'''$ are Bernouli random numbers $\in \{0,1\}$, while $T,T'',R,R''$ are deterministic.
The average value over the stochastic variable distribution]
is denoted by $\langle \cdots \rangle$.

The total charge, reaching each drain, is determined
by summing up an infinite series arising from 
multiple reflections and transmissions.
This series contains the following pieces:
\begin{enumerate}
    \item First tunnelling factor to enter the QPC.
    \item Shortest path inside the QPC.
    \item Contribution due to multiple reflections
    from different scatterers: $(R'_{12})_i (R''_{21})_i (R'''_{12})_i (R_{21})_{i+1},\ i \in [1,2,\ldots]$, which is the same for all the contacts.
\end{enumerate}

For the charge $Q_1$ which enters the drain $D_1$ the first piece is $(T_{11})_1$,
and the second piece is $(T'_{11})_n,\ n\in[1,2,\ldots]$.  How many times a wavepacket encountered the same scatterer
is denoted by the subscript outside of the parenthesis. Therefore, we write $Q_1$ as
\begin{widetext}
\begin{equation}
\begin{split}
     Q_1 &= \frac{2e}{3} \Big[ (T_{11})_1 (T'_{11})_1 + (T_{11})_1  (R'_{12})_1 (R''_{21})_1 (R'''_{12})_1 (R_{21})_2  (T'_{11})_2 \\&+ (T_{11})_1  (R'_{12})_1 (R''_{21})_1 (R'''_{12})_1 (R_{21})_2 (R'_{12})_2 (R''_{21})_2 
 (R'''_{12})_2 (R_{21})_3 (T'_{11})_3 \\&+ (T_{11})_1  (R'_{12})_1 (R''_{21})_1 (R'''_{12})_1 (R_{21})_2 (R'_{12})_2 (R''_{21})_2 (R'''_{12})_2 (R_{21})_3  (R'_{12})_3 (R''_{21})_3 (R'''_{12})_3 (R_{21})_4  (T'_{11})_4 + \ldots  \Big].
 \end{split}
\end{equation}
\end{widetext}
We point out the following conditions (and similarly for $T''',R'''$),
\begin{equation}\label{Eq-relation}
\begin{split}
&(T'_{11})_i + (R'_{12})_i = 1,\ (T'_{11})_i (R'_{12})_i = 0,\\&(T'_{22})_i + (R'_{21})_i = 1,\ (T'_{22})_i (R'_{21})_i = 0,
 \end{split}
\end{equation}
and any $T'$ and $R'$ are uncorrelated
for distinct scattering event indices $i$. We further assume that transmission and
reflection coefficients having different inner indices become uncorrelated since they correspond to different scatterers. Thus,
\begin{equation}\label{Eq-relation}
\begin{split}
&\langle Q_1 \rangle = \frac{2e}{3} \Big[  \frac{T_{11} \langle T'_{11} \rangle}{1-\langle R'_{12} \rangle R''_{21} \langle  R'''_{12} \rangle R_{21} } \Big].
 \end{split}
\end{equation}
Using $T_{11} =1, \langle T'_{11} \rangle=2/3, \langle R'_{12} \rangle=\langle R'''_{12} \rangle=1/3, R''_{21} = R_{21} =1$, we find $\langle Q_1 \rangle =e/2$. 
The source current is $I=2e/3\tau$ and the 
transmission coefficient becomes
$t=\langle Q_1 \rangle/\tau I=3/4$ and the 
QPC conductance plateau is at 
$G_{D_1}=1/2$.
The charge reaching $D_2$ is 
$\langle Q_2 \rangle = 2e/3 - \langle Q_1 \rangle=e/6$.
\begin{figure}[H]
	\includegraphics[width=0.95\columnwidth]{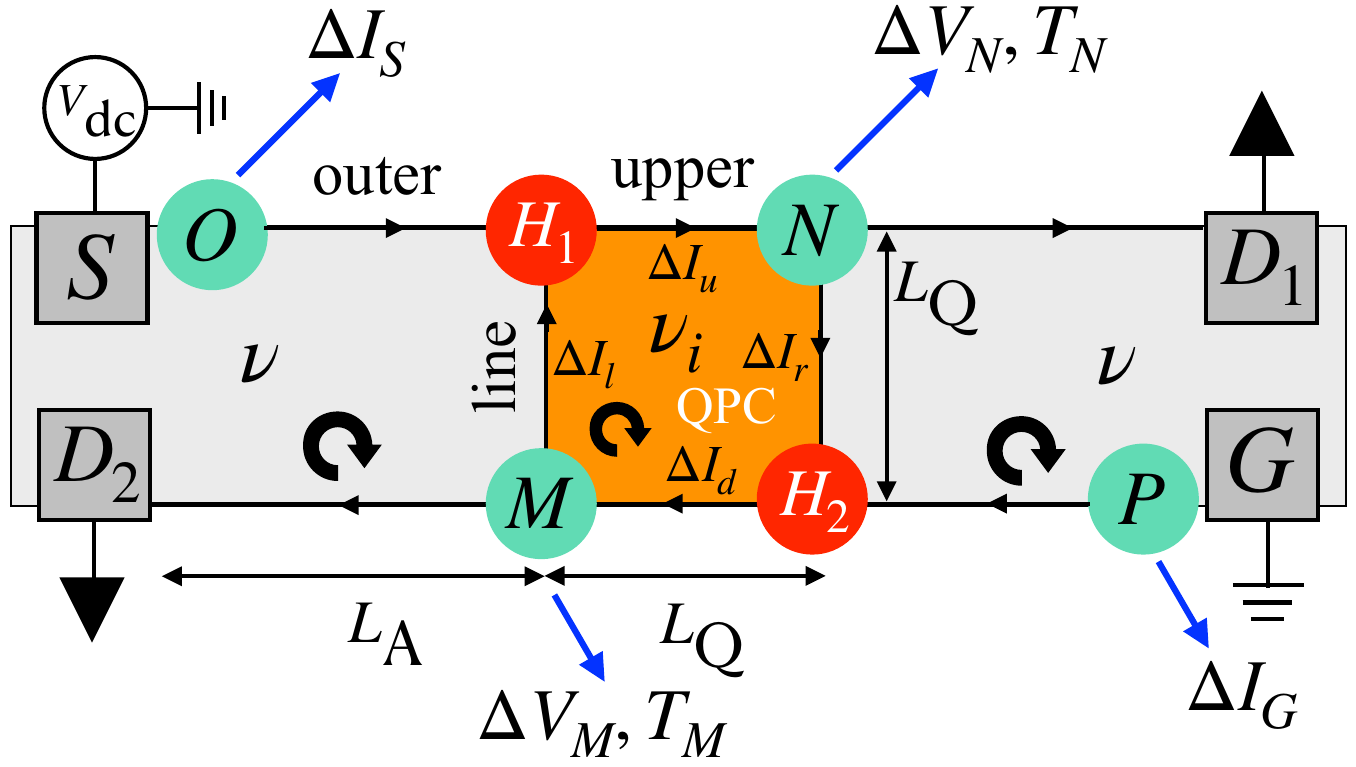}
	\caption{$1/2$ QPC conductance plateau in the
 equilibrated case: The set up is 
 similar to \cref{KFP_Fig}. The bulk filling is $\nu(=2/3)$ and the QPC filling is $\nu_i(=1)$. We have the geometric lengths: $L_{\text{A}}, L_{\text{Q}}$. We denote the segment between the vacuum and $\nu$ as  ``outer", between the vacuum and $\nu_i$ as  ``upper", and between $\nu$ and $\nu_i$ as ``line". In each segment, the arrow
 shows the fully
 equilibrated charge propagation.
 Hot spots $H_1,H_2$ (red circles) are created due to the
 voltage drops there, leading in turn to the formation of noise spots $M,N,O,P$ (green circles) \cite{PhysRevB.101.075308}.}\label{nuSmall_nuiLarge_SM}
\end{figure}

For the autocorrelation $\delta^2 Q_1=\langle Q^2_1 \rangle- \langle Q_1 \rangle^2 = \langle Q_1 \rangle(2e/3-\langle Q_1 \rangle)$,
since $\langle Q_1^2 \rangle = 2e \langle Q_1 \rangle/3$, at $D_1$ we obtain
$\delta^2 Q_1=e^2/12$.
Similarly, we write for the autocorrelation $\delta^2 Q_2=\langle Q^2_2 \rangle- \langle Q_2 \rangle^2$ at $D_2$ we obtain
$\delta^2 Q_2=e^2/12$
and the crosscorrelation $\delta^2 Q_c= \langle Q_1 Q_2 \rangle - \langle Q_1 \rangle \langle Q_2 \rangle = - \langle Q_1 \rangle \langle Q_2 \rangle$
(since $\langle Q_1 Q_2 \rangle = 0$) becomes
$\delta^2 Q_c=-e^2/12$.
The Fano factors are defined as
\begin{equation}
    F_i = \frac{\delta^2 Q_i}{e \tau I t (1-t)},\ i \in \{1,2,c\},
\end{equation}
and therefore they evaluate
to be $F_1= F_2 = -F_c = 2/3$.

\subsection{Equilibrated scenario}

For
bulk filling $\nu$ and QPC filling $\nu_i$ with $\nu<\nu_i$ (\cref{nuSmall_nuiLarge_SM}), we derive analytical expressions for the current-current correlations (CCC) (shot noise). We assume a fully equilibrated charge transport, which is ballistic, moving ``downstream" along a segment.
 We take
no bulk-leakage
\cite{PhysRevLett.123.137701, Banerjee2018, PhysRevB.99.041302,PhysRevLett.125.157702} and assume that the lead contacts are at zero temperature.

We consider \cref{nuSmall_nuiLarge_SM} and follow Refs.\ \cite{PhysRevB.101.075308,SM5by2} to write 
\begin{equation}\label{CurFluc}
	\begin{split}
		&\Delta I_S+\Delta I_l=\Delta I_u,\\&
\Delta I_u=\Delta I_1+\Delta I_r,\\&
\Delta I_G+\Delta I_r=\Delta I_d,\\&
\Delta I_d=\Delta I_2+\Delta I_l,
	\end{split}
\end{equation}
where $\Delta I_i, i \in \{S, G, u, d, r, l, 1, 2\}$ are the current fluctuations. We also write
\begin{equation}\label{VolFluc}
	\begin{split}
		&\Delta I_1 = \nu \frac{e^2}{h}\Delta V_N + \Delta I_1^{\text{th}},\\&
\Delta I_r = (\nu_i-\nu) \frac{e^2}{h}\Delta V_N + \Delta I_r^{\text{th}},\\&
\Delta I_2 = \nu \frac{e^2}{h}\Delta V_M + \Delta I_2^{\text{th}},\\&
\Delta I_l = (\nu_i-\nu) \frac{e^2}{h}\Delta V_M + \Delta I_l^{\text{th}},
	\end{split}
\end{equation}
where $\Delta V_i, i \in \{M,N\}$ are the voltage fluctuations and 
$\Delta I_i^{\text{th}}, i \in \{1,2,r,l\}$
are the thermal fluctuations. We find
\begin{equation}\label{D1Fluc}
	\begin{split}
		&\Delta I_1 = \frac{1}{(\nu-2\nu_i)}\Big[ (\nu-\nu_i)
  (\Delta I_G + \Delta I_1^{\text{th}} - \Delta I_2^{\text{th}})\\&\qquad
  - \nu (\Delta I_l^{\text{th}} - \Delta I_r^{\text{th}}) -\nu_i \Delta I_S \Big],\\&
  \Delta I_2 = \frac{1}{(\nu-2\nu_i)}\Big[ (\nu-\nu_i)
  (\Delta I_S - \Delta I_1^{\text{th}} + \Delta I_2^{\text{th}})\\&\qquad
  + \nu (\Delta I_l^{\text{th}} - \Delta I_r^{\text{th}}) -\nu_i \Delta I_G \Big].
	\end{split}
\end{equation}
We use the local Johnson-Nyquist relations for thermal noise,
\begin{equation}
	\begin{split}
		&\ \ \ \ \  \langle (\Delta I_l^{\text{th}})^2 \rangle = \frac{2 e^2}{h} (\nu_i-\nu) k_{\text{B}} T_M,\\& \ \ \ \ \ \langle (\Delta I_1^{\text{th}})^2 \rangle = \frac{2 e^2}{h} \nu k_{\text{B}} T_N,\\& \ \ \ \ \ \langle (\Delta I_r^{\text{th}})^2 \rangle = \frac{2 e^2}{h}
		(\nu_i-\nu)k_{\text{B}} T_N,\\& \ \ \ \ \ \langle (\Delta I_2^{\text{th}})^2 \rangle = \frac{2 e^2}{h} \nu k_{\text{B}} T_M,\\& \langle (\Delta I_i^{\text{th}} \Delta I_j^{\text{th}}) \rangle = 0,\ \text{for}\ i \neq j\ \text{and}\ i,j \in \{1,2,l,r\},
	\end{split}
\end{equation}
where $k_{\text{B}}$ is the Boltzmann constant to write 
\begin{equation}
	\begin{split}
		\delta^2 I_1 &=2\Bigg(\frac{e^2}{h}\Bigg)\frac{\nu \nu_i (\nu_i-\nu)}{(\nu-2\nu_i)^2}k_{\text{B}} (T_M+T_N) \\&+ \frac{1}{(\nu-2\nu_i)^2}\Big[\nu_i^2 \langle (\Delta I_S)^2 \rangle + (\nu-\nu_i )^2\langle (\Delta I_G)^2 \rangle\Big],
	\end{split}
\end{equation}
\begin{equation}
	\begin{split}
		\delta^2 I_2 &=2\Bigg(\frac{e^2}{h}\Bigg)\frac{\nu \nu_i (\nu_i-\nu)}{(\nu-2\nu_i)^2}k_{\text{B}} (T_M+T_N) \\&+ \frac{1}{(\nu-2\nu_i)^2}\Big[\nu_i^2 \langle (\Delta I_G)^2 \rangle + (\nu-\nu_i )^2\langle (\Delta I_S)^2 \rangle\Big],
	\end{split}
\end{equation}
and
\begin{equation}
	\begin{split}
		\delta^2 I_c &=-2\Bigg(\frac{e^2}{h}\Bigg)\frac{\nu \nu_i (\nu_i-\nu)}{(\nu-2\nu_i)^2}k_{\text{B}} (T_M+T_N) \\&+ \frac{\nu_i(\nu_i-\nu)}{(\nu-2\nu_i)^2}\Big[\langle (\Delta I_S)^2 \rangle + \langle (\Delta I_G)^2 \rangle\Big],
	\end{split}
\end{equation}
where the noise spot ($M$ and $N$) temperatures are $T_M, T_N$, respectively \cite{SM5by2,PhysRevB.101.075308}. The 
dissipated powers at the hot spots is
\begin{equation}
	\begin{split}
		P_{H_1} = P_{H_2} = \frac{e^2 V_{\text{dc}}^2}{h}\frac{\nu(\nu_i-\nu)}{2\nu_i} \Big( \frac{\nu_i}{2 \nu_i - \nu}  \Big)^2.
	\end{split}
\end{equation}
The contributions
$\langle (\Delta I_G)^2 \rangle = \langle (\Delta I_S)^2$
at the noise spots $O$ and $P$ can be calculated in a similar way as shown in Ref.\ \onlinecite{SM5by2,PhysRevB.101.075308}.
The Joule heating term provides a negligible contribution to the noise spots \cite{SM5by2,PhysRevB.101.075308}.

We consider $\{ \nu,\nu_i \} = \{ 2/3,1 \}, \{ 2/3(\text{r}),1 \},$ and $ \{ 2/3(\text{r}), 1(\text{r}) \}$,
where $2/3(\text{r})$ denotes the reconstructed MacDonald edge \cite{PhysRevLett.72.2624, Wang2013} having filling factor discontinuity $\delta \nu = [-1/3,+1, -1/3, +1/3]$ (\cref{nuSmall_nuiLarge_SM}). Also
$1(\text{r})$ referes to the edge reconstruction in QPC, with the discontinuity
 $\delta \nu_i = [+1, -1/3, +1/3]$ (\cref{nuSmall_nuiLarge_SM}) \cite{Khanna2021}.
For full charge equilibration (\cref{nuSmall_nuiLarge_SM}), we have 
\begin{equation}\label{I1}
	\begin{split}
		I_1 = \frac{e^2 V_{\text{dc}}}{h} \times \frac{2}{3}\times \frac{2}{3}\times \sum_{i=0}^{\infty}\Big( \frac{1}{3^2} \Big)^i=\frac{e^2 V_{\text{dc}}}{2h}
	\end{split}
\end{equation}
for each of the $\{ \nu,\nu_i \}$ choices, which leads to the 
transmission $t=3/4$ and $G_{D_1}=1/2$. 

Three distinct thermal equilibration regimes
are considered (taking $l_{\text{eq}}^{\text{th}}$ as
the thermal equilibration length):
(i) each segment of the device is thermally
unequilibrated leading to
$L_{\text{Q}} \ll L_{\text{A}} \ll l_{\text{eq}}^{\text{th}}$ (no thermal equilibration); (ii)
only the QPC segment is thermally
unequilibrated while all other 
segments are thermally equilibrated leading to
$L_{\text{Q}} \ll l_{\text{eq}}^{\text{th}} \ll L_{\text{A}}$ (hybrid thermal equilibration); (iii) or each segment is thermally equilibrated leading to
$l_{\text{eq}}^{\text{th}} \ll L_{\text{Q}} \ll L_{\text{A}}$ (full thermal equilibration).
Ref.\ \onlinecite{Kumar2022} also studied the thermally unequilibrated case for a single edge.
We note that only the upstream modes from
$H_1$($H_2$) contribute to $O$($P$) in the
no thermal equilibration regime, as the downstream modes
become fully thermally
isolated from the upstream modes and have zero temperature.

Below we compute the CCC values for
$\{ \nu,\nu_i \} = \{ 2/3,1 \}$ in the no thermal equilibration regime.
Energy conservation at $H_1, M$ leads to \cite{SM5by2,PhysRevB.101.075308} :
\begin{equation}
    \begin{split}
        &J_S+J_l+P_{H_1}=J_u,\\&J_d=J_l+J_2,
    \end{split}
\end{equation}
where the heat current $J_i$ flows along the segment
$i\in\{S, l, u, d, 2\}$. We write
\begin{equation}
    \begin{split}
        &J_S=-\frac{\kappa}{2}T^2_{H_1}, J_l=\frac{\kappa}{2}(2T^2_M-T^2_{H_1}),\\&
        J_u=\frac{\kappa}{2}T^2_{H_1},J_d=\frac{\kappa}{2}T^2_{H_2},J_2=\frac{\kappa}{2}T^2_{M},
    \end{split}
\end{equation}
where $\kappa=\frac{\pi^2 k^2_{\text{B}}}{3h}$ and 
$T_{H_1}=T_{H_2}$ and $T_M=T_N$ are the temperatures at $H_1, H_2, M, N$, respectively. We find 
\begin{equation}
    \begin{split}
        &T_M=T_N=\frac{3\sqrt{4} eV_{\text{dc}}}{4\sqrt{15} \pi k_{\text{B}}},
        T_{H_1}=T_{H_2}=\frac{3\sqrt{2} eV_{\text{dc}}}{4\sqrt{5} \pi k_{\text{B}}}.
    \end{split}
\end{equation}
The contribution from the noise spots $O$ and $P$  
is evaluated
to \cite{SM5by2,PhysRevB.101.075308}
\begin{equation}
    \begin{split}
        \langle (\Delta I_S)^2 \rangle = \langle (\Delta I_G)^2 \rangle \approx 0.0472 \frac{e^3 V_{\text{dc}}}{h}
    \end{split}
\end{equation}
leading to $F_1=F_2\approx 0.36, F_c \approx -0.17$.

In a similar manner, for 
$\{ \nu,\nu_i \} = \{ 2/3(\text{r}),1 \}$ we find 
\begin{equation}
    \begin{split}
        &T_M=T_N=\frac{ 3eV_{\text{dc}}}{16 \pi k_{\text{B}}},\\&\langle (\Delta I_S)^2 \rangle = \langle (\Delta I_G)^2 \rangle \approx 0.048 \frac{e^3 V_{\text{dc}}}{h},
    \end{split}
\end{equation}
leading to $F_1=F_2\approx 0.23, F_c \approx -0.04$.
For 
$\{ \nu,\nu_i \} = \{ 2/3(\text{r}),1(\text{r}) \}$ we find 
\begin{equation}
    \begin{split}
        &T_M=T_N=\frac{3\sqrt{5} eV_{\text{dc}}}{4\sqrt{36} \pi k_{\text{B}}},\\&\langle (\Delta I_S)^2 \rangle = \langle (\Delta I_G)^2 \rangle \approx 0.065 \frac{e^3 V_{\text{dc}}}{h},
    \end{split}
\end{equation}
leading to $F_1=F_2\approx 0.33$ and $F_c \approx -0.07$.
In a similar manner, we have diffusive 
heat transport \cite{PhysRevB.101.075308} in the outer segment for hybrid and full thermal equilibration,
leading to a length
dependent CCC. For all the choices of $\{ \nu,\nu_i \}$, we find
\begin{equation}
    \begin{split}
        &F_1=F_2\approx 0.13+0.26\sqrt{L_{\text{A}}/l_{\text{eq}}^{\text{th}}},\\&
        F_c\approx 0.07-0.26\sqrt{L_{\text{A}}/l_{\text{eq}}^{\text{th}}}.
    \end{split}
\end{equation}

\section{The noisy $5e^2/9h$ QPC plateau}\label{5by9QPC_PRB}

\begin{figure}[!t]
	\includegraphics[width=0.95\columnwidth]{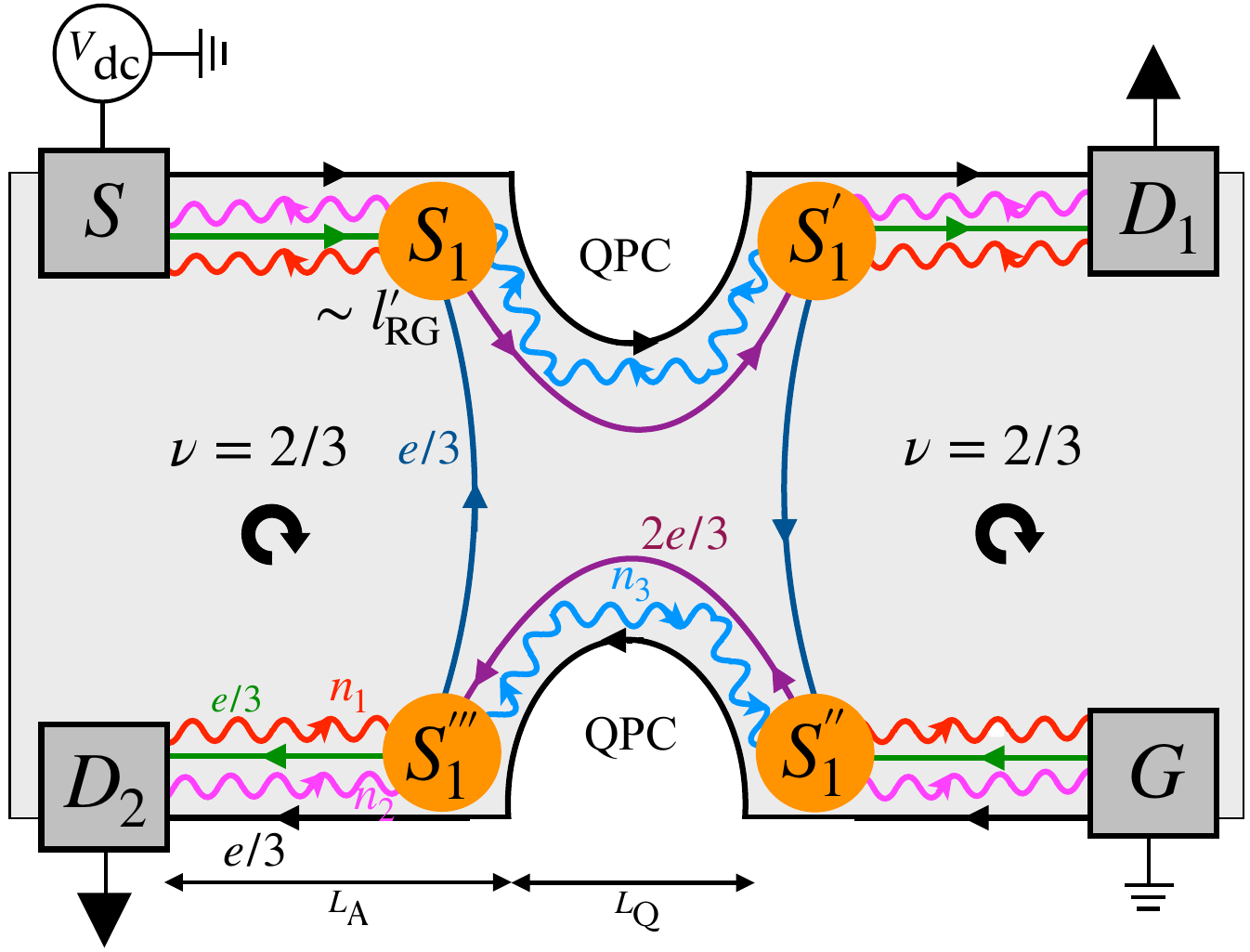}
	\caption{The stochastic scenario for the $5/9$ QPC conductance plateau where the
 normalized modes emanate from the contacts. 
 The set up is similar to \cref{KFP_Fig}. The reconstructed MacDonald edge
	structure consists of counter-propagating $e/3\ (\text{innermost}), e, e/3$ and $e/3\ (\text{outermost})$ charge modes (from bulk to edge) \cite{Wang2013}. At the 
 QPC one $e/3$ mode (outermost) transmits fully and the other $e/3$ mode (innermost) backscatters fully. The rest of the modes are renormalized to
a $2e/3$ charge mode and $n_3$ neutral mode (counter-propagating) at the KFP RG fixed point \cite{Kane1994} (KFP region). Between a contact and QPC we have 
$ n_1, n_2$ neutral modes and two $e/3$ charge modes (counter-propagating) \cite{Wang2013} (WMG region). Near QPC, we have regions of size $l'_{\text{RG}}$ (orange circles). We define these regions by 
using density kernel matrices $S_1,S_1',S_1'', S_1'''$ \cite{PhysRevB.52.R17040} constructed out of the
transmission and reflection coefficients.
We note that $S_1'$ and $S_1'''$ describe
stochastic processes while $S_1$ and $S_1''$
describe deterministic ones.}\label{WMG_Fig}
\end{figure}

Here, we show how the $5e^2/9h$ QPC conductance plateau 
can appear
only in the unequilibrated regime and compute
shot noise at this plateau. We consider only the stochastic scenario.
We note that if we consider the equilibration
then it becomes exactly the same scenario of 
\cref{1by2QPC_PRB}.

The bare edge structure contains the 
counter-propagating $e/3$ (``innermost"), $e$, $e/3$ and $e/3$ (``outermost")
charge modes as we go from the bulk towards the edge; this is the reconstructed MacDonald edge \cite{Wang2013} (\cref{WMG_Fig}).
We label the $e/3, e, e/3, e/3$ charge modes (from bulk to edge in the contacts) as $``2", ``1", ``3", ``4"$, respectively. 
At the QPC, we consider the case when the outermost $e/3$ mode is fully transmitted, the innermost $e/3$ mode is fully backscattered, and
the remaining $e$ and $e/3$ modes are renormalized
to the KFP RG fixed point \cite{Kane1994},
giving rise to counter-propagating $2e/3$ 
charge mode and a $n_3$ neutral mode. This regime will be called the KFP region.
Similarly, we will call the WMG regime the case where the 
renormalized (at the WMG RG fixed point \cite{Wang2013}) charge modes $e/3, e/3$ counter-propagate to the
$n_1, n_2$ neutral modes between each contact and the QPC.

Following \cref{KFP_stoch}, let us then assume that the renormalized modes emerge from the
contacts (\cref{WMG_Fig}), and that we have incoherent scattering of wavepackets in
each region of size $l'_{\text{RG}}$ near the QPC.
These processes are quantified by the density kernel matrices $S_1,S'_1,S''_1,S'''_1$ \cite{PhysRevB.52.R17040}. 
We point out that $S_1'$ and $S_1'''$ are responsible for stochastic processes where incoming charge is split between the outgoing modes,
while $S_1$ and $S_1''$ bear the deterministic processes, where incoming charge has only a single outgoing mode to proceed to.
Explicitly, we write
\begin{equation}
\begin{split}
&S_1=
\begin{pmatrix}
T_{11}  &  R_{21} & R_{31} \\
R_{12}  &  T_{22}  &  R_{32} \\
R_{13}  &  R_{23}  &  T_{33} 
\end{pmatrix}
=S_1'',\\&
S_1'=
\begin{pmatrix}
\langle T'_{11} \rangle & \langle R'_{21} \rangle & \langle R'_{31} \rangle\\
\langle R'_{12} \rangle & \langle T'_{22} \rangle & \langle R'_{32} \rangle\\
\langle R'_{13} \rangle & \langle R'_{23} \rangle & \langle T'_{33} \rangle
\end{pmatrix}
=S_1'''.
\end{split}
\end{equation}
We note that the outermost $e/3$
mode does not take part in the noise generation
but does enter the calculation of the transmission 
coefficient. In a similar manner as before, we write
\begin{equation}\label{Eq-relation}
\begin{split}
&\langle Q_1 \rangle = \frac{e}{3} \Big[  \frac{T_{11}  \langle T'_{11} \rangle}{1-\langle R'_{12} \rangle R''_{21} \langle  R'''_{12} \rangle R_{21} } \Big].
 \end{split}
\end{equation}
We use $T_{11}=1, \langle T'_{11} \rangle=1/2, \langle R'_{12} \rangle=\langle R'''_{12} \rangle=1/2, R''_{21} = R_{21} =1$
to find $\langle Q_1 \rangle =2e/9$ and
$\langle Q_2 \rangle =e/9$. 
The source current is $I=2e/3\tau$ and the 
transmission coefficient is then 
$t=(\langle Q_1 \rangle + e/3)/\tau I=5/6$ and the 
QPC conductance plateau is at 
$G_{D_1}=5/9$.

The current autocorrelation at $D_1$ is $\delta^2 Q_1=\langle Q^2_1 \rangle- \langle Q_1 \rangle^2 = \langle Q_1 \rangle(e/3-\langle Q_1 \rangle)$.
Since $\langle Q_1^2 \rangle = e \langle Q_1 \rangle/3$, we obtain
$\delta^2 Q_1=2e^2/81$.
Similarly, for the autocorrelation $\delta^2 Q_2=\langle Q^2_2 \rangle- \langle Q_2 \rangle^2$ at $D_2$ we obtain
$\delta^2 Q_2=2e^2/81$, while the crosscorrelation $\delta^2 Q_c= \langle Q_1 Q_2 \rangle - \langle Q_1 \rangle \langle Q_2 \rangle = - \langle Q_1 \rangle \langle Q_2 \rangle$
(since $\langle Q_1 Q_2 \rangle = 0$) becomes
$\delta^2 Q_c=-2e^2/81$.
The Fano factors are
\begin{equation}
    F_i = \frac{\delta^2 Q_i}{e \tau I t (1-t)},\ i \in \{1,2,c\},
\end{equation}
and therefore we find them
to be $F_1= F_2 = -F_c \approx 0.266$. 

\section{The noisy $e^2/3h$ QPC plateau}\label{1by3QPC_PRB}

Here, we show how $e^2/3h$ QPC conductance plateau 
(experimentally discovered in Ref.\ \onlinecite{Mahalu2009},
following earlier theoretical works in Refs.\ \onlinecite{Sabo2017,Park2021})
can appear
both in the unequilibrated (only the stochastic scenario) and equilibrated regimes. We compute
the shot noise at this plateau and show that different inequalities 
hold among the autocorrelations and crosscorrelation
for the different scenarios.

\subsection{Unequilibrated scenario}

\begin{figure}[]
	\includegraphics[width=0.95\columnwidth]{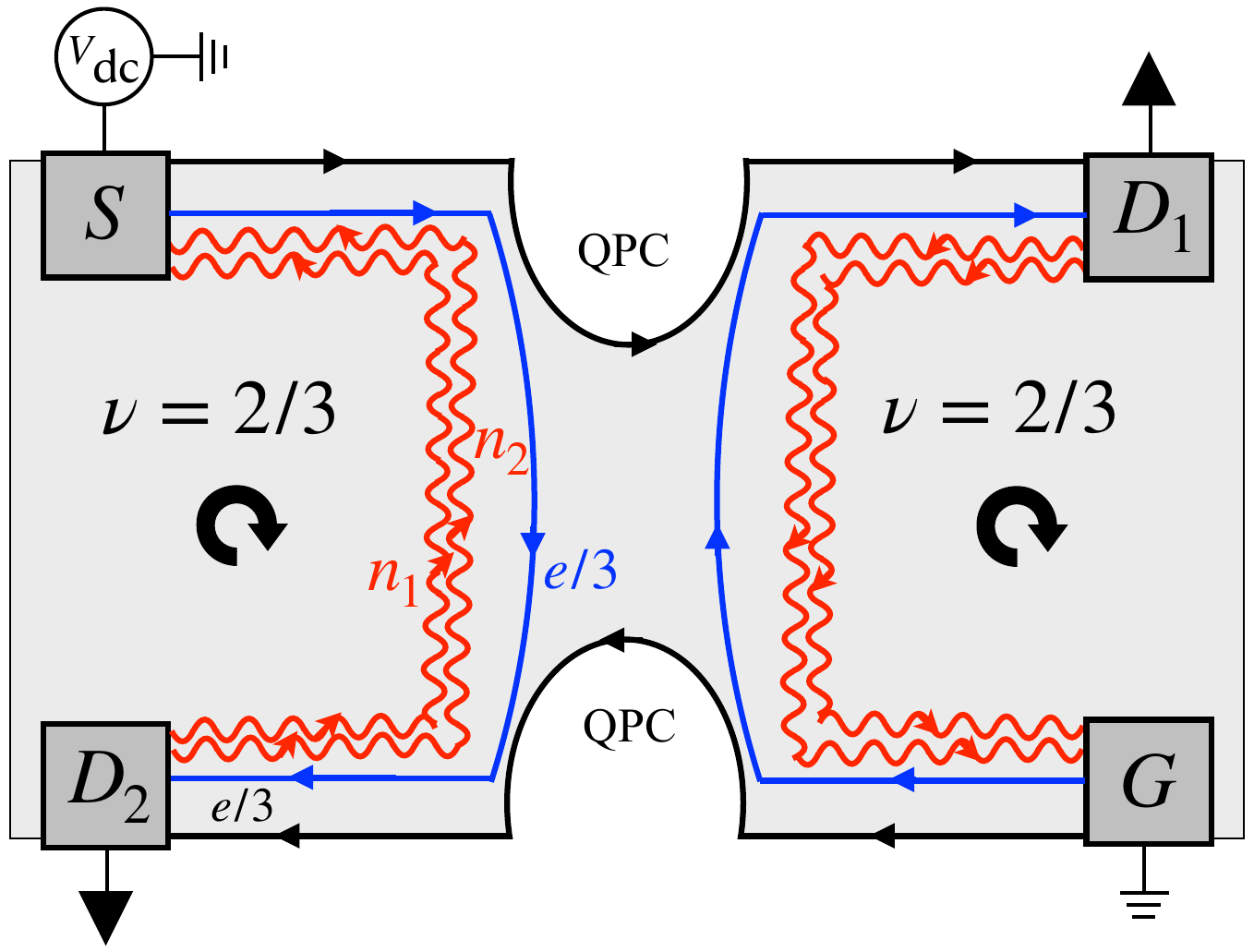}
	\caption{The unequilibrated stochastic scenario for the $1/3$ QPC conductance plateau, where the
 renormalized modes emanate from the contacts. We refer to \cref{KFP_Fig} for the
 geometry. Here the modes (going from the bulk
 towards the edge) are $n_1,n_2,e/3\ (\text{inner}),e/3\ (\text{outer})$. We denote by $n_1, n_2$ the neutral modes, which we draw as the wiggly red. At the QPC, the inner $e/3$ charge mode gets backscattered completely while the outer $e/3$ charge mode gets transmitted fully.}\label{WMG_RGFP_Fig}
\end{figure}

We consider the 
renormalized (at the WMG RG fixed point \cite{Wang2013}) reconstructed MacDonald edge structure, which consists of 
the $n_1,n_2, e/3\ (\text{inner})$ and $e/3\ (\text{outer})$
modes (from bulk to edge), where $n_1, n_2$ denote the neutral modes (\cref{WMG_RGFP_Fig}).
A plateau is observed at transmission $t=1/2$, leading to $(G_{D_1}e^2)/h$ QPC conductance plateau, where $G_{D_1}=t I \tau/e$ and $I$ is the source current, when the inner $e/3$ charge mode is fully backscattered and the outer $e/3$ charge mode is fully transmitted \cite{Mahalu2009}. The neutral modes counter-propagate with respect to the charge modes and are fully backscattered at the QPC. We calculate the shot noise below following Refs.\ \onlinecite{Sabo2017,Park2021}.

The source biases the two $e/3$ charge modes. We assume that $N$
quasiparticles each having charge $e/3$ emanate from $S$ into each charge mode in a time interval $\tau$. 
Near the lower right side of QPC, upon equilibration $N_1=N/2$ quasiparticles populate each mode and neutralons are created. These neutralons move to the upper right side of QPC via the neutral modes and randomly decay into quasihole-quasiparticle 
pairs in the adjacent charge modes \cite{Park2021}. This decay process is stochastic and lead to $N_1$ and $(N-N_1)$ electronic excitations in the inner and outer modes, respectively, which reach $D_1$ and $D_2$ and generate nonzero dc shot noises but zero dc current. 
In a similar manner, near
the upper left side of QPC equilibration takes place, and after equilibration $N_2=N/2$ quasiparticles present in both modes. This process also creates neutral
excitations which move to the lower left side of QPC and stochastically decay into $(N-N_2)$ and $N_2$ electronic excitations in the inner and outer modes, respectively,
which reach $D_2$ and $D_1$ and generate shot noise.

We introduce random variables $a,b$, assuming the values $\pm 1$ with equal probability, which characterize the neutral decay processes near upper right side and lower left side of QPC, respectively. We have $\langle a_m^{p/q} \rangle = \langle b_m^{p/q}
\rangle = 0$. Here $p$ stands for the inner $e/3$ charge mode and $q$ stands for the outer $e/3$ charge mode. We note that $a$ and $b$ are mutually uncorrelated and use the following properties.
\begin{equation}
\begin{split}
  &\langle a_m^{p}a_n^{q} \rangle = \langle b_m^{p}b_n^{q} \rangle = \delta_{m,n}\delta_{p,q},\\& \langle a_m^{p} a_n^{q} \rangle = \langle b_m^{p} b_n^{q} \rangle = -\delta_{m,n}\ \text{for}\ p \neq q.
 \end{split}
\end{equation}
The charges $Q_1$ and $Q_2$ reaching at drains $D_1$ and $D_2$ during time $\tau$ are thus
\begin{equation}
\begin{split}
 &Q_1 = eN_1/3 + eN_1/3 + e/3 \sum_{i=1}^{N_1} a_i^{p} + e/3 \sum_{j=1}^{N_2} b_j^{q},\\&Q_2 = eN_2/3 + eN_2/3 + e/3 \sum_{k=1}^{N-N_1} a_k^{q} + e/3 \sum_{l=1}^{N-N_2} b_l^{p}.  
 \end{split}
\end{equation}
The total current is
\begin{equation}
\begin{split}
I = \frac{\langle Q_1 \rangle + \langle Q_2 \rangle}{\tau} = \frac{2eN}{3\tau},
 \end{split}
\end{equation}
the transmission through QPC is
\begin{equation}
\begin{split}
t=\frac{\langle Q_1 \rangle}{\langle Q_1 \rangle + \langle Q_2 \rangle} = \frac{1}{2},
 \end{split}
\end{equation}
and the QPC conductance plateau is at $ G_{D_1} = 1/3$.
The autocorrelation at $D_1$ is then
\begin{equation}
\begin{split}
 \delta^2 Q_1 &= \langle (Q_1 - \langle Q_1 \rangle)^2 \rangle\\&=\frac{e^2}{9} \Bigg\langle \Bigg(\sum_{i=1}^{N_1} a_i^{p} + \sum_{j=1}^{N_2} b_j^{q}\Bigg) \Bigg(\sum_{m=1}^{N_1} a_m^{p} + \sum_{n=1}^{N_2} b_n^{q}\Bigg)\Bigg\rangle\\&=\frac{e^2N}{9}. 
 \end{split}
\end{equation}
Similarly we obtain for the autocorrelation at $D_2$, $\delta^2 Q_2 = \langle (Q_2 - \langle Q_2 \rangle)^2 \rangle = e^2N/9$, whereas the crosscorrelation is
\begin{equation}
\begin{split}
 \delta^2 Q_c &= \langle (Q_1 - \langle Q_1 \rangle) (Q_2 - \langle Q_2 \rangle) \rangle\\&=\frac{e^2}{9} \Bigg\langle \Bigg(\sum_{i=1}^{N_1} a_i^{p} + \sum_{j=1}^{N_2} b_j^{q}\Bigg) \Bigg(\sum_{k=1}^{(N-N_1)} a_k^{q} \\&+ \sum_{l=1}^{(N-N_2)} b_l^{p}\Bigg)\Bigg\rangle
 \\& = - \frac{e^2N}{9}.
 \end{split}
\end{equation}
Finally, the Fano factor for autocorrelation at $D_1$ is \cite{Sabo2017,Park2021}
\begin{equation}
\begin{split}
 F_1=\frac{\delta^2 Q_1}{e \tau I t (1-t)}=2/3,
 \end{split}
\end{equation}
and similarly we have $F_2 = -F_c = 2/3$.

\subsection{Equilibrated scenario}

\begin{figure}[!t]
	\includegraphics[width=0.95\columnwidth]{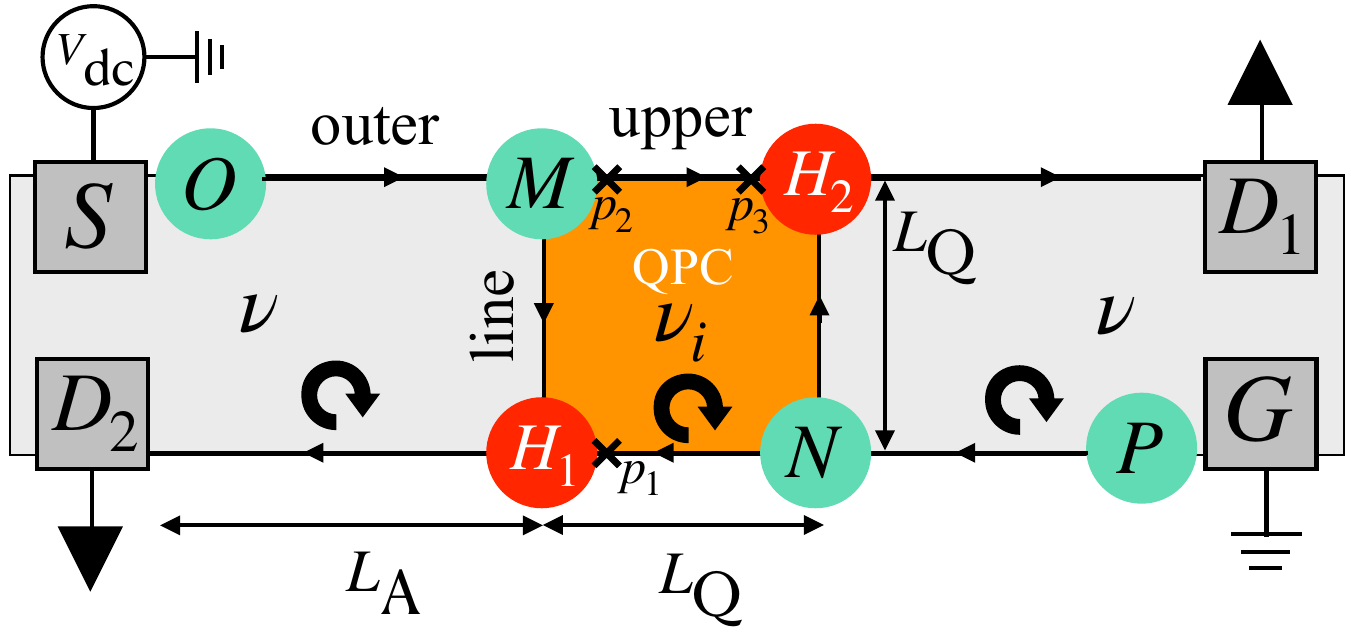}
	\caption{The equilibration scenario for the $1/3$ QPC conductance plateau. The setup is similar to \cref{nuSmall_nuiLarge_SM}, but now we have $\nu>\nu_i$. Hot spots $H_1,H_2$ (red circles) occur due to the voltage drops and noise spots $M,N,O,P$ (green circles) \cite{PhysRevB.101.075308} result thereby.}\label{nuLarge_nuiSmall_SM}
\end{figure}

We derive the general expressions for the CCC (shot noise) in a QPC for the
bulk filling $\nu$ and QPC filling $\nu_i$ when $\nu>\nu_i$ (\cref{nuLarge_nuiSmall_SM}). We assume that the charge is fully equilibrated, hence charge transport is ballistic, moving downstream along each segment of the setup.
We call the direction opposite to charge flow (upstream) antiballistic.
Thereafter, we compute the values of CCC for 
specific choices of $\{\nu,\nu_i\}$ and for different
thermal equilibration regimes.
 We assume that there is 
no bulk-leakage
\cite{PhysRevLett.123.137701, Banerjee2018, PhysRevB.99.041302,PhysRevLett.125.157702}.
We study the scenarios employing two pictures for the thermal equilibration, which were introduced in Ref.\ \onlinecite{SMshort} and Ref.\ \onlinecite{Park2024}, respectively. The experimental situation may be expected to be intermediate between these two pictures.

\subsubsection{The picture of Ref.\ \onlinecite{SMshort}}

Here the picture is that each of the modes creates noise
by producing particle-hole pairs at the noise spots
due to very short charge equilibration length. These processes lead to local heat
equilibration by transfering heat
energy among the modes. Thereby, all the modes
acquire a common temperature, leading to CCC 
which is independent of length (c.f.\ \cref{nuLarge_nuiSmall_SM}). 

We follow Refs.\ \onlinecite{PhysRevB.101.075308,SM5by2} and recompute the expressions of $\delta^2 I_1$ (autocorrelation in drain $D_1$), $\delta^2 I_2$ (autocorrelation in drain $D_2$) and $\delta^2 I_c$  (crosscorrelation) (\cref{nuLarge_nuiSmall_SM}) in the same spirit of \cref{1by2QPC_PRB}.

We write
\begin{equation}
	\begin{split}
		\Delta I_1 &= \Delta I_u + \Delta I_r,\\ \Delta I_2 &= \Delta I_d + \Delta I_l,\\
  \Delta I_S &= \Delta I_u + \Delta I_l,\\
  \Delta I_G &= \Delta I_d + \Delta I_r,
	\end{split}
\end{equation}
and
\begin{equation}
	\begin{split}
		& \Delta I_l = (\nu-\nu_i) \frac{e^2}{h} \Delta V_M + \Delta I_l^{\text{th}},\\& \Delta I_u = \nu_i \frac{e^2}{h} \Delta V_M + \Delta I_u^{\text{th}},\\&\Delta I_r = (\nu-\nu_i) \frac{e^2}{h} \Delta V_N + \Delta I_r^{\text{th}},\\& \Delta I_d = \nu_i \frac{e^2}{h} \Delta V_N + \Delta I_r^{\text{th}}.
	\end{split}
\end{equation}
Thereby, we find
\begin{equation}
	\begin{split}
		\Delta I_1 &= \nu_i \frac{e^2}{h} \Delta V_M + \Delta I_u^{\text{th}}+(\nu-\nu_i) \frac{e^2}{h} \Delta V_N + \Delta I_r^{\text{th}}\\&=\frac{\nu_i}{\nu}\Delta I_S +
		\frac{(\nu-\nu_i)}{\nu}\Delta I_G \\&+ \frac{(\nu-\nu_i)}{\nu}(\Delta I_u^{\text{th}} - \Delta I_d^{\text{th}})+\frac{\nu_i}{\nu}(\Delta I_r^{\text{th}} - \Delta I_l^{\text{th}})
	\end{split}
\end{equation}
and
\begin{equation}
	\begin{split}
		\Delta I_2 &=\frac{\nu_i}{\nu}\Delta I_G + \frac{(\nu-\nu_i)}{\nu}\Delta I_S \\&+
		\frac{(\nu-\nu_i)}{\nu}(\Delta I_d^{\text{th}} - \Delta I_u^{\text{th}})+\frac{\nu_i}{\nu}(\Delta I_l^{\text{th}} - \Delta I_r^{\text{th}}).
	\end{split}
\end{equation}
Employing the local Johnson-Nyquist relations for thermal noise,
\begin{equation}
	\begin{split}
		&\ \ \ \ \  \langle (\Delta I_l^{\text{th}})^2 \rangle = \frac{2 e^2}{h} (\nu-\nu_i) k_{\text{B}} T_M,\\& \ \ \ \ \ \langle (\Delta I_u^{\text{th}})^2 \rangle = \frac{2 e^2}{h} \nu_i k_{\text{B}} T_M,\\& \ \ \ \ \ \langle (\Delta I_r^{\text{th}})^2 \rangle = \frac{2 e^2}{h}
		(\nu-\nu_i)k_{\text{B}} T_N,\\& \ \ \ \ \ \langle (\Delta I_d^{\text{th}})^2 \rangle = \frac{2 e^2}{h} \nu_i k_{\text{B}} T_N,\\& \langle (\Delta I_i^{\text{th}} \Delta I_j^{\text{th}}) \rangle = 0,\ \text{for}\ i \neq j\ \text{and}\ i,j \in \{l,u,r,d\},
	\end{split}
\end{equation}
we find the correlations to be
\begin{equation}
	\begin{split}
		\delta^2 I_1 &=2\Bigg(\frac{e^2}{h}\Bigg)\frac{\nu_i}{\nu}(\nu-\nu_i)k_{\text{B}} (T_M+T_N) \\&+ \frac{1}{\nu^2}\Big[\nu_i^2 \langle (\Delta I_S)^2 \rangle + (\nu-\nu_i )^2\langle (\Delta I_G)^2 \rangle\Big],
	\end{split}
\end{equation}
\begin{equation}
	\begin{split}
		\delta^2 I_2 &=2\Bigg(\frac{e^2}{h}\Bigg)\frac{\nu_i}{\nu}(\nu-\nu_i)k_{\text{B}} (T_M+T_N) \\&+
		\frac{1}{\nu^2}\Big[\nu_i^2 \langle (\Delta I_G)^2 \rangle + (\nu-\nu_i )^2\langle (\Delta I_S)^2 \rangle\Big],
	\end{split}
\end{equation}
and
\begin{equation}
	\begin{split}
		\delta^2 I_c &=-2\Bigg(\frac{e^2}{h}\Bigg)\frac{\nu_i}{\nu}(\nu-\nu_i)k_{\text{B}} (T_M+T_N) \\&+ \frac{\nu_i(\nu-\nu_i)}{\nu^2}\Big[\langle (\Delta I_G)^2 \rangle + \langle (\Delta I_S)^2 \rangle\Big],
	\end{split}
\end{equation}
where $T_M, T_N$ are, respectively, the temperatures at the noise spots $M$ and $N$, which are found by solving
self-consistent equilibration equations and 
considering energy conservation at each edge mode junction \cite{SM5by2,PhysRevB.101.075308}. 
We note that the 
dissipated powers at the hot spots take the form
\begin{equation}
	\begin{split}
		P_{H_1} = P_{H_2} = \frac{e^2 V_{\text{dc}}^2}{h}\frac{\nu_i(\nu-\nu_i)}{2\nu}.
	\end{split}
\end{equation}
The contributions
$\langle (\Delta I_G)^2 \rangle = \langle (\Delta I_S)^2$
at the noise spots $O$ and $P$ are computed by evaluating integrals, as shown in Ref.\ \onlinecite{SM5by2,PhysRevB.101.075308}, assuming that no voltage drops take place along the outer segment 
and the lead contacts remain at zero temperature..

Here we consider $\{ \nu,\nu_i \} = \{ 2/3,1/3 \},$ and $ \{ 2/3(\text{r}), 1/3) \}$.
As charge is fully equilibrated in each segment of the QPC set up 
(\cref{nuLarge_nuiSmall_SM}), we have $I_1=e^2 V_{\text{dc}}/(3h)$
for each of the $\{ \nu,\nu_i \}$ choices here, leading to
transmission $t=1/2$ and $G_{D_1}=1/3$ \cite{PhysRevB.101.075308}. 

For no thermal equilibration (considered also in Ref.\ \onlinecite{Kumar2022} for a single edge segment), we have only ballistic and
antiballistic heat transports in any segment of the QPC set up, leading to a constant CCC. 
Similarly, by following our assumptions as before, only upstream modes will
contribute at $O, P$.
For 
$\{ \nu,\nu_i \} = \{ 2/3,1/3 \}$ we find 
\begin{equation}
    \begin{split}
        &T_M=T_N=\frac{ eV_{\text{dc}}}{\sqrt{5} \pi k_{\text{B}}},\\&\langle (\Delta I_S)^2 \rangle = \langle (\Delta I_G)^2 \rangle \approx 0.044 \frac{e^3 V_{\text{dc}}}{h},
    \end{split}
\end{equation}
leading to $F_1=F_2\approx 0.35, F_c \approx -0.22$.
For 
$\{ \nu,\nu_i \} = \{ 2/3(\text{r}),1/3 \}$ we find
\begin{equation}
    \begin{split}
        &T_M=T_N=\frac{3 eV_{\text{dc}}}{4\sqrt{6} \pi k_{\text{B}}},
        \\
        &\langle (\Delta I_S)^2 \rangle = \langle (\Delta I_G)^2 \rangle \approx 0.06 \frac{e^3 V_{\text{dc}}}{h},
    \end{split}
\end{equation}
leading to $F_1=F_2\approx 0.28, F_c \approx -0.1$.
Similarly, for hybrid and full thermal equilibration, we have diffusive 
heat transport in the outer segment and 
ballistic and
antiballistic heat transports in the line and
the upper segments of the QPC set up, leading to length
dependent CCC. For each choice of $\{ \nu,\nu_i \}$, we find
\cite{PhysRevB.101.075308}
\begin{equation}
    \begin{split}
        &F_1=F_2\approx 0.09+0.3\sqrt{L_{\text{A}}/l_{\text{eq}}^{\text{th}}},\\&F_c\approx 0.09-0.3\sqrt{L_{\text{A}}/l_{\text{eq}}^{\text{th}}}.
    \end{split}
\end{equation}

\subsubsection{The picture of Ref.\ \onlinecite{Park2024}}

Here, not all the modes reach a common temperature at the noise spot. Rather, the mode(s) going around the QPC slowly
give up heat to the modes going around the bulk regions (c.f.\ \cref{nuLarge_nuiSmall_SM}).
We calculate the noise in the no thermal
equilibration regime by following the assumption
of Ref.\ \onlinecite{Park2024}.
We consider a scenario as in \cref{nuLarge_nuiSmall_SM}.
Ref.\ \onlinecite{Park2024} 
assumed that none of the modes acquire a 
common temperature at the noise spot. In addition,
the mode with filling $\nu_{i}$ slowly gives heat to the mode with filling $\nu$ while travelling.

We write the differential
equation for heat transport along the AB segment as
\cite{PhysRevB.101.075308}
\begin{equation}
\begin{split}
    &\partial_x \begin{pmatrix}
  T^2(x) \\
  T_{i}^2(x) \\
\end{pmatrix} = \frac{1}{l^{\text{th}}_{\text{eq}}} \begin{pmatrix}
-1 & 1\\
-1 & 1
\end{pmatrix} \begin{pmatrix}
  T^2(x) \\
  T_{i}^2(x) \\
\end{pmatrix} -\frac{2P}{\kappa}\delta(x-x_{\text{hot}}),
\end{split}
\end{equation}
where $x_{\text{hot}}$ is the hotspot location and $P$ is the power dissipated there. $T$ and $T_{i}$ are the temperature profiles
for filling $\nu$ and $\nu_{i}$,
respectively.
We find the diffusive temperature profiles for the AB segment as
\begin{equation}
\begin{split}  &T^2(x)=c_1+c_2x,\\&T_{i}^2(x)=c_1+c_2(x+l^{\text{th}}_{\text{eq}}).
\end{split}
\end{equation}
In addition, $T^2_{i}$ jumps by $2P/\kappa$ upon crossing the hotspot location $x=L_Q$ when going from the segment $NH_1$ to $H_1M$. Now one needs to impose the appropriate boundary conditions, solve for the temperature profiles, and thereby find the temperature $T_{i}(x=0)$ at the noise spot $M$.

One boundary condition is $T(x=0)=0$, which gives $c_1=0$. Now, 
we consider three points $p_1, p_2, p_3$ as in the figure.
From the symmetry of the system, we conclude that 
$T^2_{i}(x = p_1) = T^2_{i}(x = p_3)$. 
Also we note that
$T^2_{i}(x = p_2) = T^2_{i}(x = p_3)$, since only the mode with filling $\nu_{i}$ is travelling from $M$ to $H_2$. Hence, we write 
\begin{equation}
\begin{split}
    &T^2_{i}(x = p_2) = T^2_{i}(x = p_1),\\& c_2 l^{\text{th}}_{\text{eq}}= c_2 (L_Q+l^{\text{th}}_{\text{eq}})-\frac{2P}{\kappa},\\& c_2 = \frac{2P}{\kappa L_Q}.
\end{split}
\end{equation}
Thus, we find 
$T_{i}(x=0)=\sqrt{2Pl^{\text{th}}_{\text{eq}}/\kappa L_Q}$, leading to shot noise $\sim \sqrt{l^{\text{th}}_{\text{eq}}/L_Q}$.

\section{Summary and outlook}\label{Summ}
In this work we have considered the $\nu=2/3$ FQH state in a QPC geometry. We have studied 
both the bare and reconstructed edge structures that are consistent with the
bulk-boundary correspondence. For each of these, we have 
considered the 
steady state of the edge modes to be either unequilibrated or
equilibrated. 
In the unequilibrated regime, 
two different paradigms of RG fixed points have been proposed earlier, namely KFP \cite{Kane1994} and WMG \cite{Wang2013}.
Recent experiments \cite{PhysRevLett.126.216803, Melcer2022, Kumar2022, Srivastav2022} have established that the thermal equilibration length
is an order of magnitude large than the charge equilibration length.
These findings imply that  
different degrees of thermal equilibration 
are possible, along with full equilibration as far as charge is concerned. In this study we have found that there are three possible
QPC-generated two-terminal conductance plateaus,
$e^2/2h, 5e^2/9h,$ and $e^2/3h$; there are several distinct scenarios (discussed above) that lead to the first and last  conductance values.
Experimentally, the $e^2/2h$ plateau has been reported 
recently \cite{Manfra2022,Hirayama2023} and the 
$e^2/3h$ plateau was discovered earlier \cite{Mahalu2009}.
We predict that if one finds the $e^2/3h$ plateau, then another
possible plateau can exist at $5e^2/9h$ if the edge modes are
in the unequilibrated regime. 
To figure out in which regime we are in, we can use
\emph{electrical shot noise}, namely
autocorrelations and crosscorrelation, at these plateaus.
We have identified that distinct mechanisms are responsible
for the existence of shot noise in different scenarios, leading to different inequalities between them.
Thus, our comprehensive study 
provides a classification of 
the steady state of the edge modes based on shot noise.
Our proposal can be extended to other filling fractions \cite{SMAD}
and can be
tested in the experiments with present day 
technology.

\begin{acknowledgements}
We thank Christian Glattli, Udit Khanna,
Michael J.\ Manfra, Alexander D.\ Mirlin,
Jinhong Park, Christian 
Sp\aa{}nsl\"att, and Kun Yang for useful discussions.
We thank Christian Glattli also for sharing his unpublished data.
S.M.\ was supported by the Weizmann Institute of Science, Israel Deans fellowship through Feinberg
Graduate School, as well as the Raymond Beverly Sackler Center for Computational Molecular and Material Science at Tel Aviv University. 
A.D.\ was supported by DFG
MI 658/10-2 and DFG RO 2247/11-1. A.D.\ also thanks the Israel Planning and budgeting committee (PBC) and the
Weizmann Institute of Science, the Dean of Faculty fellowship, and the Koshland Foundation for financial support. A.D.\ thanks IISER Tirupati start-up grant for support.
Y.G.\ acknowledges support by the DFG Grant MI 658/10-2, by the German-Israeli Foundation Grant
I-1505-303.10/2019, by the ISF grant, by the DFG
Grant RO 2247/11-1, and by the Minerva Foundation.
Y.G.\ is the incumbent of InfoSys Chair.
M.G.\ has been supported by the Israel Science Foundation (ISF) and the Directorate for Defense
Research and Development (DDR\&D) Grant No. 3427/21, the ISF grant No. 1113/23, and the US-Israel Binational Science Foundation (BSF) Grant No. 2020072. 
\end{acknowledgements}

\bibliography{bibfile}	
\end{document}